\documentclass{article}

\usepackage{xcolor}
\usepackage[nonatbib]{ml4eng_2020}
\usepackage[utf8]{inputenc} 
\usepackage[T1]{fontenc}    
\usepackage{booktabs}
\usepackage{multirow}
\usepackage[flushleft]{threeparttable}
\usepackage{graphicx}
\usepackage{caption}
\usepackage{subcaption}
\usepackage{alphalph}
\usepackage{algorithm,algpseudocode}

\usepackage{algorithm}
\usepackage{amsmath}
\usepackage{enumitem}
\setlist{
  noitemsep,
  listparindent=\parindent,
  parsep=0pt,
}
\makeatletter
\def\algbackskip{\hskip\dimexpr-\algorithmicindent+\labelsep}
\def\LState{\State \algbackskip}%
\makeatother

\newcommand\MyHead[2]{%
  \multicolumn{1}{l}{\parbox{#1}{\centering #2}}
}

\usepackage{setspace}

\usepackage{titlesec}

\titlespacing{\section}{0pt}{12pt}{6pt}
\titlespacing{\subsection}{0pt}{6pt}{6pt}
\titlespacing{\subsubsection}{0pt}{4pt}{4pt}

\title{A Sequential Modelling Approach for Indoor Temperature Prediction and Heating Control in Smart Buildings}

\author{%
  Yongchao Huang \\
  University of Oxford\\
  \texttt{yongchao.huang@cs.ox.ac.uk} \\
  \And
  Hugh Miles \\
  Atamate \emph{Ltd.} \\
  \texttt{hugh.miles@atamate.com} \\
  \AND
  Pengfei Zhang \\
  Facebook, London \\
  \texttt{zhangp1@fb.com} \\
}

\begin{document}

\maketitle

\begin{abstract}

  The rising availability of large volume data, along with increasing computing power, has enabled a wide application of statistical Machine Learning (ML) algorithms in the domains of Cyber-Physical Systems (CPS), Internet of Things (IoT) and Smart Building Networks (SBN). This paper proposes a learning-based framework for sequentially applying the data-driven statistical methods to predict indoor temperature and yields an algorithm for controlling building heating system accordingly. This framework consists of a two-stage modelling effort: in the first stage, an univariate time series model (AR) was employed to predict ambient conditions; together with other control variables, they served as the input features for a second stage modelling where an multivariate ML model (XGBoost) was deployed. The models were trained with real world data from building sensor network measurements, and used to predict future temperature trajectories. Experimental results demonstrate the effectiveness of the modelling approach and control algorithm, and reveal the promising potential of the mixed data-driven approach in smart building applications. By making wise use of IoT sensory data and ML algorithms, this work contributes to efficient energy management and sustainability in smart buildings.

\end{abstract}


\section{Introduction}
Internet of Things (IoT) technologies introduce a new generation of smart building solutions \cite{intel, Chrysi}. It embeds network platforms and advanced data analytics with the Building Management System (BMS) which integrates lighting, heating, ventilation, and air conditioning (HVAC), safety, and security, turning sensory data into to actionable decisions. IoT has a wide spectrum of applications such as smart facilities maintenance and efficient energy management \cite{Kelly}, an example is using temperature, CO$_2$ and humidity sensors to provide the feedback for HVAC systems, or using motion detectors for lighting control. An IoT ecosystem normally consists of a comprehensive sensor network for data collection, communication, and networking, and data consuming solutions. Machine Learning (ML) plays an increasingly significant role in IoT decision-making (AI-IoT). The goal of building intelligence is to build an automated, intelligent, and decentralised decision-making system that monitors building service status and interprets them as machine-readable commands. In this setting, sensor network sits in the centre; along with data consuming platforms, it can yield significant savings for operations and maintenance, without degrading occupants’ comfort level. Take for example the heating and cooling systems, which can account for up to 60\% of a building’s overall use \cite{Morgan_Stanley}, by efficiently learning use patterns and managing on-off controls, it can reduce energy usage by 30\% as reported. This serves as the main motivation for heating control. 

Traditional building management solutions use complex building dynamics to achieve control objectives; more recently emerging techniques such as big data analytics \cite{Wenzhuo}, ML \cite{Luis,Manar,Jan} and deep reinforcement learning techniques have been deployed \cite{Ruoxi}. A typical scenario of application is to meet a target temperature at a pre-set time. In this pursuit, we aim to decide the best time to switch on the radiator, and to switch off thereafter without degrading comfort. These are the two optimisation tasks: the first helps avoid wasting utility in case the valve is turned on too early; the second gives obvious energy-saving bonus. This is particularly the case when one finishes work and asks BMS to heat up a room before actually arriving home. This paper focuses on the first task, while the solutions provided can be safely extended to the second. The challenge is to accurately predict the warm-up time, and we introduced a sequential, data-driven modelling approach to tackle it.

\section{Related work}
Applying ML \cite{Luis,Manar} and big data analytics \cite{Wenzhuo} to IoT (particularly smart buildings) has been an increasingly hot area in recent years. Some previous work on modelling and controlling building energy systems were summarised, for example, in \cite{Harish, Xiwang, Hrvoje, Saleh}. Mathematical (structured) modelling (also termed the 'white box' approach) has long been used in simulating natural phenomenon such as heat diffusion, it’s well constructed based on physical laws of thermal dynamics, and normally comes in the form of partial differential equations with some stochastic characteristics (\textit{e.g.} the state-space models describing a dynamical system). For example, \cite{Degurunnehalage} describes a mathematical model for heating modelling in residential buildings, their results demonstrate the effectiveness of the model-based heating control strategy. Non-linear systems can also be designed by purpose for special buildings such as \cite{Zhi}. Mathematical models have intuitive interpretability and mature solvers. However, they may not be as flexible as statistical models which are learning-based, data-driven, and independent of the physical system. In statistical modelling (also termed 'black box' approach), each signal is treated as an input feature, it’s thus easier to take control over the variables fed in. The models are based on statistical principles which analyse the relation between input features and the output variable. Since statistical models are physics-independent, they can be universally used to model various phenomena by learning from data. The disadvantage, however, is also obvious: they depend on the partial info contained in the data used to train it. Following the data allows fast adjustments but also introduces bias. Results from a data-driven approach can yield good accuracy, but care should be taken on generalisation. \cite{Saleh2} emphasizes the importance of specific tuning for each ML model used in the prediction of buildings heating and cooling loads. \cite{Saleh} gives a review on the use of several ML methods in forecasting building energy performance, and concluded that all models provide reasonable accuracy given large samples and fine tuned hyper-parameters. A compromise could be combining both the 'white box' and 'black box' approaches to form a semi-structured modelling approach (the 'grey box' or hybrid approach), in which known laws are represented by deterministic terms, and unknown effects or uncertainties can be embedded via learning-based models (\textit{e.g.} a neural network architecture). This bridges the gap between physical and statistical modelling: it enables prior info to be embedded, the building dynamics captured, and gives a reliable description of the uncertainties \cite{Henrik}, and results in a simpler models as compared to pure physics based models \cite{Kumar}.

 While the power of the data driven approach has been proved in many smart building settings (\textit{e.g.} HVAC control \cite{Donald, Tianshu}), it also has some drawbacks with estimation \cite{Niels} and forecasting. Common ML models, \textit{Random Forest} for example, rely on instantaneous, cross-sectional ambient info to make inference about a response. In most situations, however, collection of the input features may not be in a timing manner. Even if the data can be collected and processed in near real-time, there is inevitably a delay in forecasting, \textit{i.e.} the algorithm can only make inference about the current state where surrounding information is provided. Therefore, these methods are more suitable for inferencing a currently unobservable response variable rather than making swift predictions. The game changes when future information becomes available. Future information itself can be predicted in a structured way if it follows some dynamics, and in sensor network analysis particularly, most signals are time series sequences and there are mature techniques well suited for forecasting. For example, there are analytical forms such as an autoregressive process, and architectures such as an \textit{LSTM} network for analysing sequential data. This modelling approach serves as a pathway to obtaining future information about the quantity of interest. Once the future information is estimated to certain accuracy, they can be fed into common ML models to make predictions for the response variable. This motivates the two-stage forecasting framework.

\section{A Two-stage Sequential Modelling Approach}
\begin{figure}[h!]
    \centering
    \includegraphics[width=0.75\textwidth]{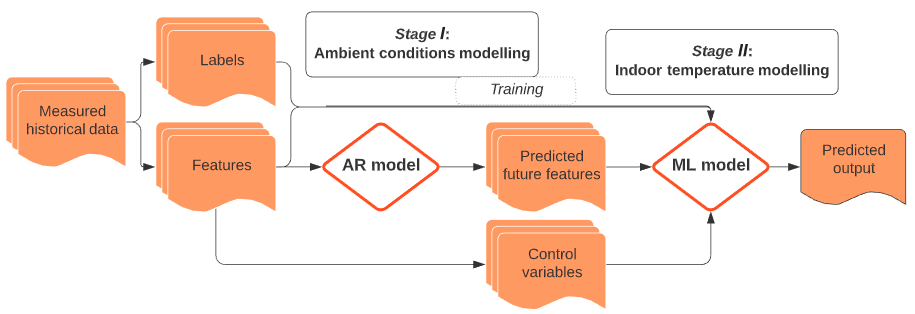}
    \caption{The two-stage sequential modelling procedure}
    \label{fig:two_stage_modelling}
\end{figure}

We proposed a two-stage modelling procedure for household indoor temperature prediction and heating control. The framework was shown in Fig.\ref{fig:two_stage_modelling}. It's a mixed data-driven method which combines input control and learning. The sensory data were classified into two categories: control and non-control variables. In the first stage, each non-control variable time series was treated as a univariate process and time series models (\textit{e.g.} AR) were employed to make predictions for a future period; in the second stage, the control variables (with future values set according to some control rules such as on/off), together with the non-control variable predictions, were fed into a multivariate ML model (pre-trained using historically measured data) to yield final predictions for the output quantity (indoor temperature). In later contexts, the first practice was labelled ‘ambient conditions modelling’, while the second was termed ‘indoor temperature modelling’.

\vspace{2.5mm}
\setlength{\intextsep}{8pt}

\section{Building Management System (BMS) and Sensor Network}
A BMS is a comprehensive network enabling data flow between different functional components and making decisions based on pre-defined rules and algorithms. The front ends are sensors, which monitor and detect changes in environmental and device variables such as temperature, air quality, valves and occupancy to optimise a building's operation services like HVAC over time. Data was collected by a sensor unit which accommodates an array of sensors such as temperature, lighting, fan, radiator meters, Passive Infrared motion (PIR) sensor, rain detector, and relative humidity (RH) reader. It flows into data containers and processors where data storage, aggregation, transformation, inference and prediction happen. The challenge is to deploy the system in a real-time environment at scale. For example, data may be collected overtime in an augmented manner so that time-variant adjustments can be made. Retraining a predictive model could be either done online (streaming data) or offline (batch data). For large learning tasks, transfer or incremental learning could be leveraged via a de-centralised data consuming architecture. These put requirements on both the physical and software sides. For example, sensors may need to talk to each other wirelessly, data needs to be transported either locally (\textit{e.g.} offline training) or on cloud, and processors may need to communicate for collaboration purpose. Terminal devices such as water tanks, controllers (\textit{e.g}. radiators) and air conditioners need to be plugged in BMS to enable an state-action feedback loop for services control. In addition, extensibility is desired so that new sensors, gateways (5G highway for example) and devices, can be flexibly added, upgraded and integrated into the existing BMS. 

\begin{figure}[h!]
    \centering
    \begin{subfigure}{0.55\textwidth}
        \includegraphics[height=4.5cm]{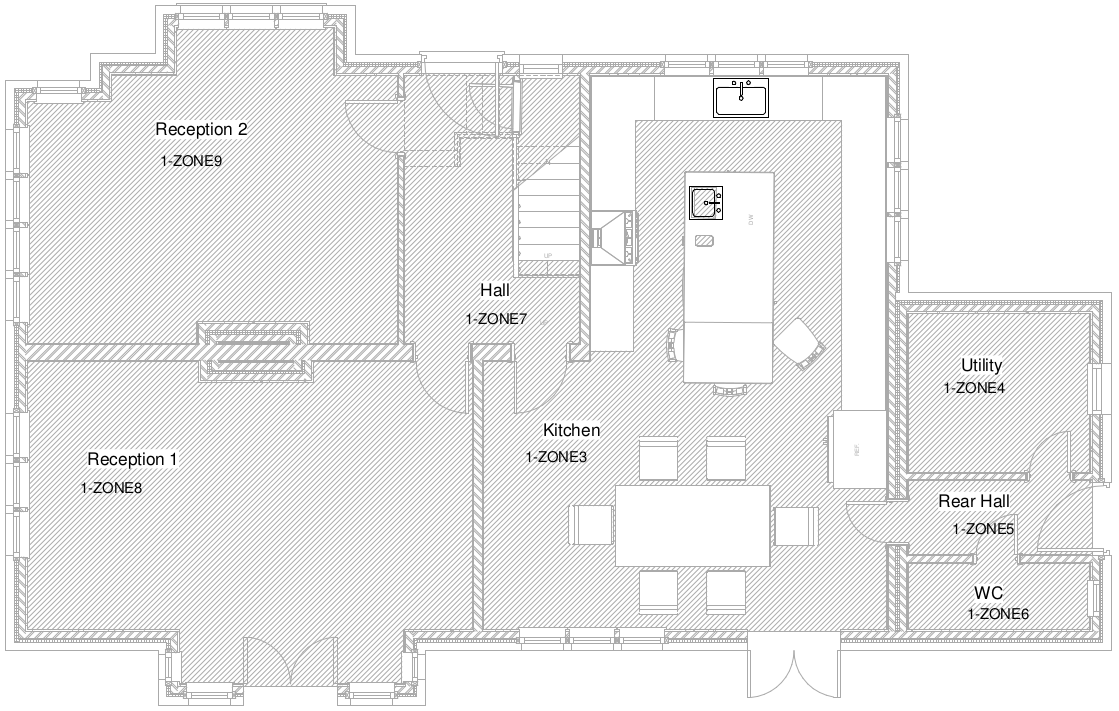}
    \end{subfigure}
    \begin{subfigure}{0.44\textwidth}
    \includegraphics[height=4.5cm]{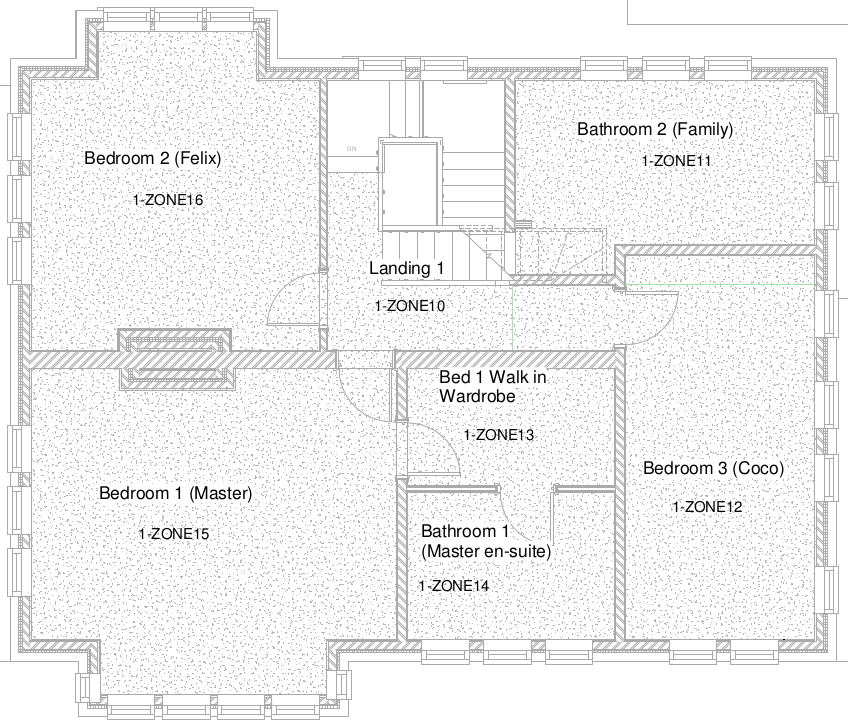}
    \end{subfigure}
    \caption{Building floor layout (left: Floor 0, right: Floor 1)}
    \label{fig:building_floor_layout}
\end{figure}

Sensors play a key role in engineering scenes such as Cyber-Physical Systems (CPS), IoT and Smart Building Networks (SBN). Thanks to the increasing capacity of sensor network, large volume of data becomes available for analysis. Making wise use of sensory data could potentially improve energy use and achieve better building management. In this research, multiple functional sensors were installed in a house environment (Fig.\ref{fig:building_floor_layout}) to monitor quantities such as temperature, lighting, fan, heating, occupancy, rain and humidity statuses. In total there were 54 sensors used. The details of the set of sensors deployed were shown in Table.\ref{tab:sensor_categories} and Table.\ref{tab:physical_mapping_of_sensor_layout}. These sensors together formed a network. 

\vspace{1mm}
\setlength{\intextsep}{12pt}
\begin{table}[!htb]
  \begin{threeparttable}
  \scriptsize
  \caption{Sensor categories}
  \label{tab:sensor_categories}
  \begin{tabular}{lccl}
    \toprule  
    \textbf{Sensor type} & \MyHead{3.5cm}{\textbf{Sensor code}\\(functionality abbreviations)} & \textbf{Sensory data type} & \textbf{Example instance(s)} \\
    \midrule
    Temperature & TMP & numeric & \emph{0-1-TMP1} \\
    Lighting & AMB/LTC & Boolean(0/1) & \emph{0-1-AMB1/0-1-LTC1} \\
    Fan & FAN &  Boolean(0/1) & \emph{1-14-FAN1} \\
    Radiator & HTC/HTV &  Boolean(0/1) & \emph{1-12-HTC1/1-1-HTV1} \\
    Occupancy & PIR/EPIR/OPC &  Boolean(0/1) & \emph{1-13-PIR1/0-1-EPIR1/1-8-OPC1} \\
    Rain & RAIN & Boolean(0/1) & \emph{0-1-RAIN2} \\
   Relative humidity & HYGR & numeric & \emph{0-1-HYGR1} \\
   \bottomrule
  \end{tabular}
  \begin{tablenotes}
      \scriptsize
      \item $\ast$Sensor naming convention: the first number implies type of space, \emph{\emph{i.e.}} 0 refers to outside space and 1 is house. The second number gives the zone area in the building (see the floor layout plan). The strings are sensor code, followed by sensor number.
    \end{tablenotes}
  \end{threeparttable}
\end{table}
\setlength{\intextsep}{4pt}
\begin{table}[!htb]
  \scriptsize
  \caption{Physical mapping of sensor layout}
  \label{tab:physical_mapping_of_sensor_layout}
  \begin{tabular}{ll}
    \toprule  
    \textbf{Zone area of sensor location} & \textbf{Sensors deployed} \\
    \midrule
    0-1 Drive & \emph{0-1-AMB1, 0-1-EPIR1, 0-1-HYGR1, 0-1-LTC1, 0-1-RAIN2, 0-1-TMP1} \\
    0-2 Garden & \emph{0-2-AMB1, 0-2-HYGR1, 0-2-TMP1} \\
    1-1 Basement & \emph{1-1-HTV1, 1-1-HTV3} \\
    1-3 Kitchen (Floor 0) & \emph{1-3-AMB1, 1-3-HTC1, 1-3-HYGR1, 1-3-LTC1, 1-3-OPC1, 1-3-PIR1} \\
    1-7 Hall (Floor 0) & \emph{1-7-AMB1, 1-7-OPC1, 1-7-PIR1, 1-7-TMP1} \\
    1-8 Reception1 (Floor 0) & \emph{1-8-AMB1, 1-8-HTC1, 1-8-OPC1, 1-8-PIR1, 1-8-TMP1} \\
   1-9 Sitting Room (Floor 0) & \emph{1-9-AMB1, 1-9-HTC1, 1-9-HYGR1, 1-9-LTC1, 1-9-OPC1, 1-9-PIR1} \\
   1-12 Bedroom3 (Floor 1) & \emph{1-12-AMB1, 1-12-HTC1, 1-12-HYGR1, 1-12-LTC1} \\
   1-13 Walk in Wardrobe (Floor 1) & \emph{1-13-AMB1, 1-13-HTV1, 1-13-HTV4, 1-13-LTC1, 1-13-OPC1, 1-13-PIR1} \\
   1-14 Ensuite Bathroom (Floor 1) & \emph{1-14-AMB1, 1-14-FAN1, 1-14-HTC1, 1-14-HYGR1, 1-14-OPC1, 1-14-PIR1, 1-14-TMP1} \\
   1-15 Master Bedroom (Floor 1) & \emph{1-15-AMB1, 1-15-LTC1, 1-15-OPC1, 1-15-PIR1, 1-15-TMP1} \\
   \bottomrule
  \end{tabular}
\end{table}
\setlength{\textfloatsep}{0pt}
\setlength{\intextsep}{4pt}

\vspace{5mm}
\setlength{\textfloatsep}{8pt}

\section{Experiments}

\subsection{Data Pre-processing}
The data was sliced from a large dataset spanning over the whole year of 2019 in the house; only December data (‘2019-12-01 00:00:00’ to ‘2019-12-30 14:39:00’) was chosen because it demonstrates the use of heating. The raw data consists of 44640 datapoints and 54 sensor signals. Prior to formal modelling, data pre-processing was performed to clean data and perform exploratory data analysis.

\begin{figure}[h!]
    \centering
    \includegraphics[width=0.65\columnwidth]{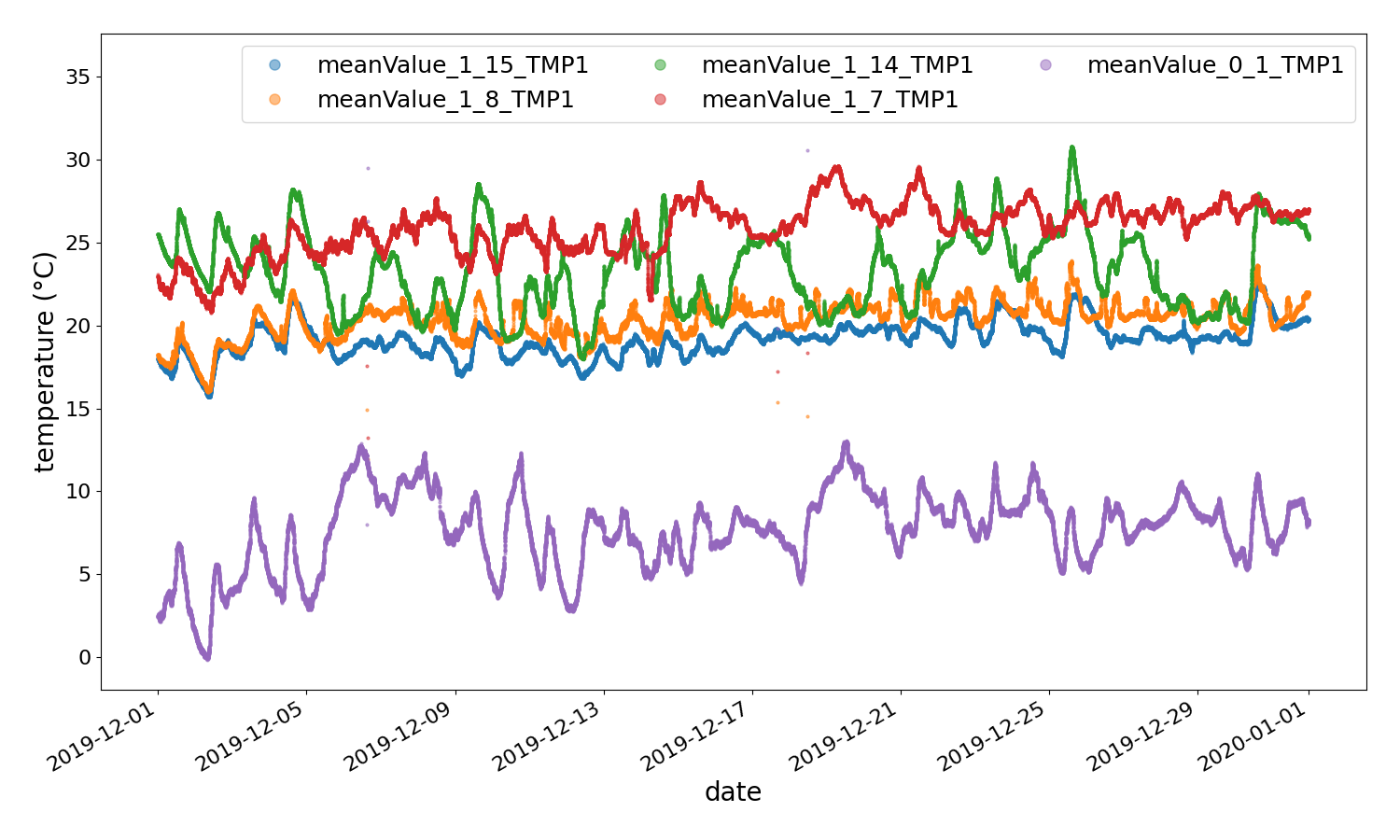}
    \caption{Raw temperature data}
    \label{fig:raw_temperature_data}
\end{figure}

In this task, we aim to predict the indoor temperature (the response variable) in the master bedroom sitting on the first floor, \textit{i.e.} 1-15-TMP1. All 54 signals were used as raw data, and a feature selection process, based on feature importance ranking, was performed. Some sensors might have different sensing frequencies; to unify different sampling rates, all data was resampled every 1 minute, which generated mean values within the sampling interval. Five raw temperature histories, demonstrated later to be most correlated with 1-15-TMP1, were shown in Fig.\ref{fig:raw_temperature_data}. It’s not surprising to find that, the temperature in Reception1 (1-8-TMP1) closely follows the target temperature (1-15-TMP1) as they are located down and upstairs. The same applies to 1-8-TMP1 and 1-14-TMP1. The Hall temperature (1-7-TMP1) deviates more as it’s an open transition area; and the outside temperature 0-1-TMP1 is affected by the weather and thus more variable.

\vspace{3mm}
\setlength{\textfloatsep}{8pt}
\setlength{\textfloatsep}{0pt}

\vspace{2mm}
Two outliers, identified based on response variable values, were excluded from the dataset, and no missing value detected. Two dimensions were added: hours and weekday information were generated from timestamp info. Weekdays were category type and encoded into binary integer features. After feature engineering, the data was split into training and test sets, with feature matrix size of (42638, 61), (2000, 61), respectively. After feature engineering, the feature space contains 53 original features, 7 weekday indicators, and one hours info column.

\begin{figure}[h!]
    \centering
    \begin{subfigure}{0.55\textwidth}
        \includegraphics[height=4.5cm]{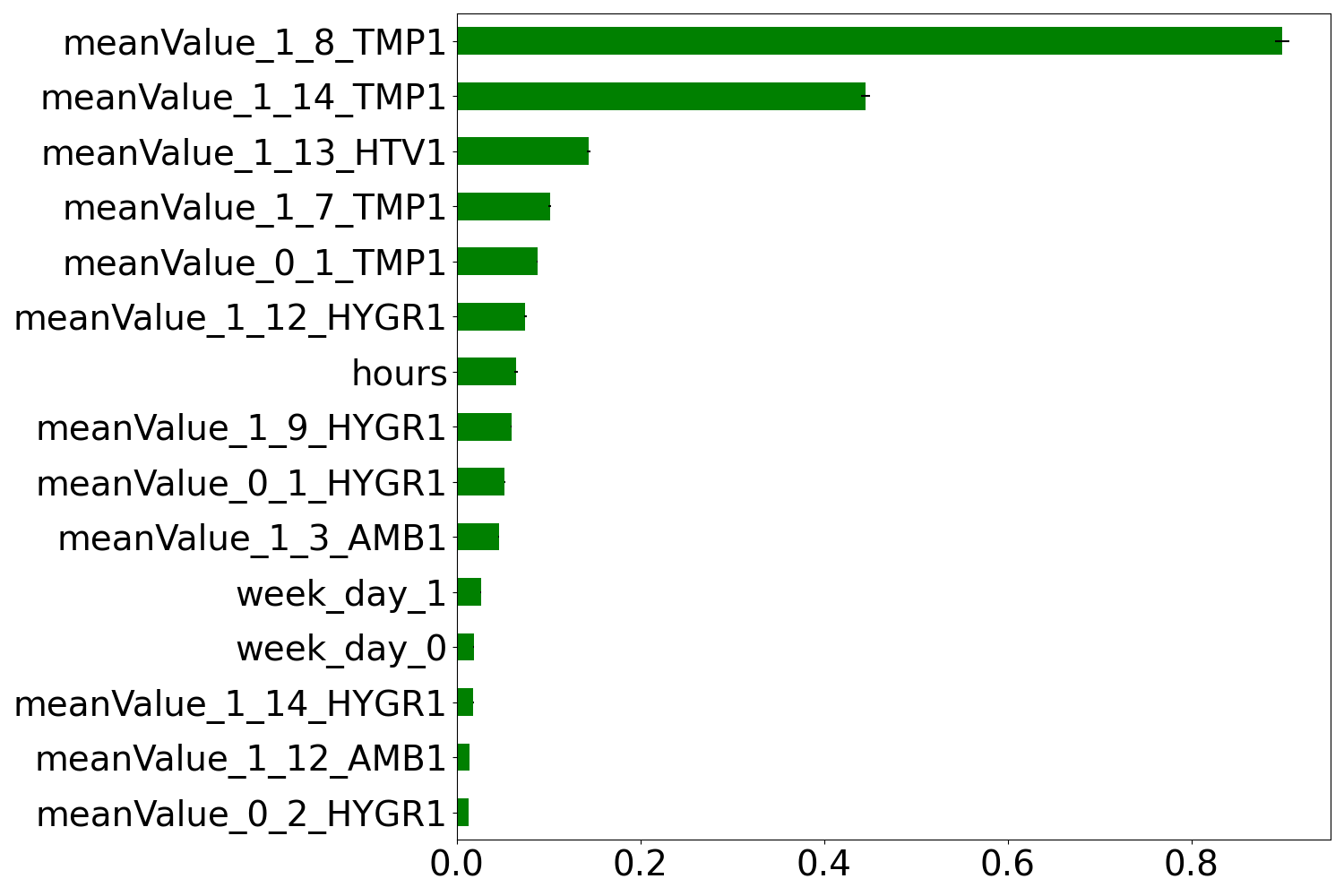}
    \end{subfigure}
    \begin{subfigure}{0.44\textwidth}
        \includegraphics[height=4.5cm]{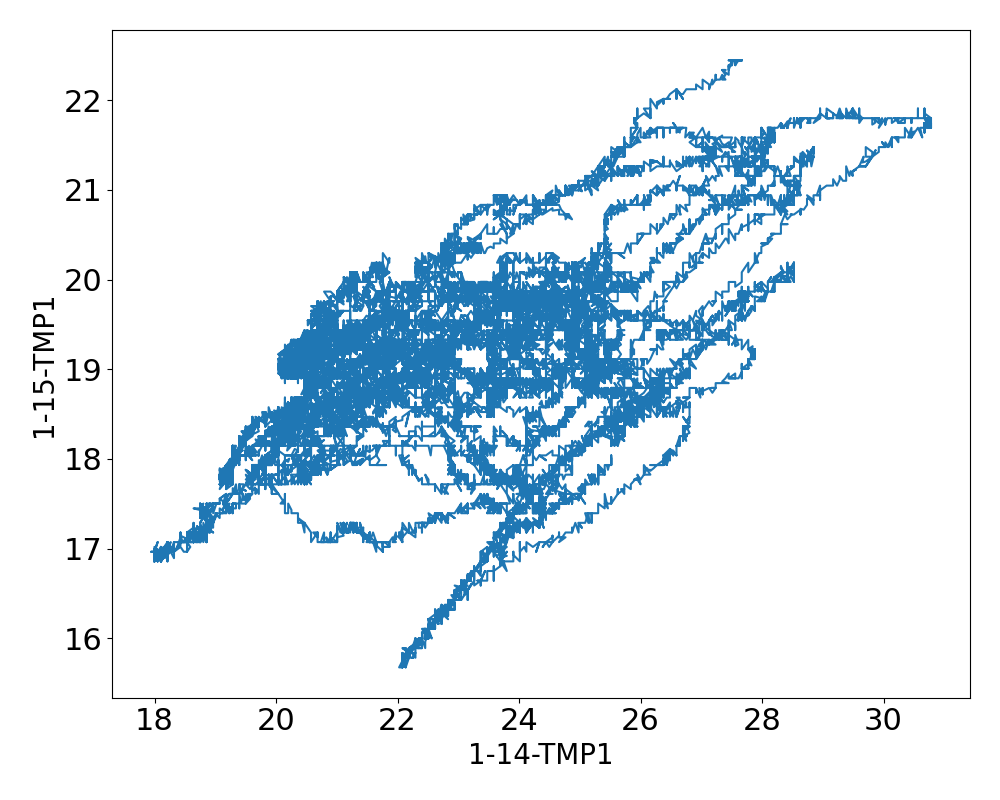}
    \end{subfigure}
    \caption{Left: top 15 RF-based permutation importance, right: correlation between 1-14-TMP1 and 1-15-TMP1}
    \label{fig:permutation_importance_and_correlation}
\end{figure}

Before modelling, a feature selection process was performed to for dimension reduction purpose. A simple \textit{Random Forest} (RF) model was trained and the resulted permutation importance was calculated to rank the 53 raw features, the top 15 ranked features are shown in Fig.\ref{fig:permutation_importance_and_correlation}. As expected, the downstairs room temperature 1-8-TMP1 is important in predicting the response variable, \textit{i.e.} the temperature in bedroom 1-15 (1-15-TMP1). Other close friends are 1-14-TMP1 and 1-13-HTV1: the next-door temperature and the status of the valve controlling underfloor heating of Room 1-15, respectively. A roughly monotonic relation between 1-14-TMP1 and 1-15-TMP1 is observed in Fig.\ref{fig:permutation_importance_and_correlation}. Besides, 1-15-TMP1 also exhibits some patterns with certain weekdays and hours in a day. Based on the importance analysis, the top 15 features, as listed in Fig.\ref{fig:permutation_importance_and_correlation}, plus the remaining weekday features, which totals 19 variables, were selected. As stated, these 19 features are classified into two categories: control and non-control variables. In a first trial, the underfloor heating valve status (1-13-HTV1) was employed as the single control variable. A second trial also included the next-door temperature 1-14-TMP1 and downstairs temperature 1-8-TMP1 as extra control variables for achieving more aggressive control. The rest 18 (16 in the second case) variables served as non-control features.
\vspace{3mm}
\setlength{\intextsep}{8pt}
\setlength{\textfloatsep}{0pt}
\subsection{Ambient Conditions Modelling}
The first step in the two-stage modelling practice is to model the non-control variables, \textit{i.e.} the ambient conditions. All the 18 features, except for weekdays and hours, need to be learned from history and predicted for a future period. Each variable trajectory was treated as a single univariate time series and modelled using autoregressive regression (AR). AR is a mature technique for modelling sequential data, the general form of an AR(p) model can be described as
\setlength{\intextsep}{4pt}
\begin{equation} \label{eq:AR}
  X_t = \mu + \alpha_1X_{t-1} + \alpha_2X_{t-2} + ... + \alpha_pX_{t-p} + \epsilon_t 
\end{equation}

\noindent where X is the time series, $\mu$ is the constant offset, and $\epsilon$ is the error term. $\alpha$s are the AR parameters to be estimated using historical realisations. For stationary signals, $\alpha$ lies in (-1,1).

\begin{figure}[h!]
\centering
\begin{subfigure}{0.9\textwidth}
\subfloat[\footnotesize{Historical and controlled/predicted values for \emph{1-13-HTV1} and \emph{1-14-TMP1}}]{\includegraphics[clip,width=1\columnwidth,height=4.6cm]{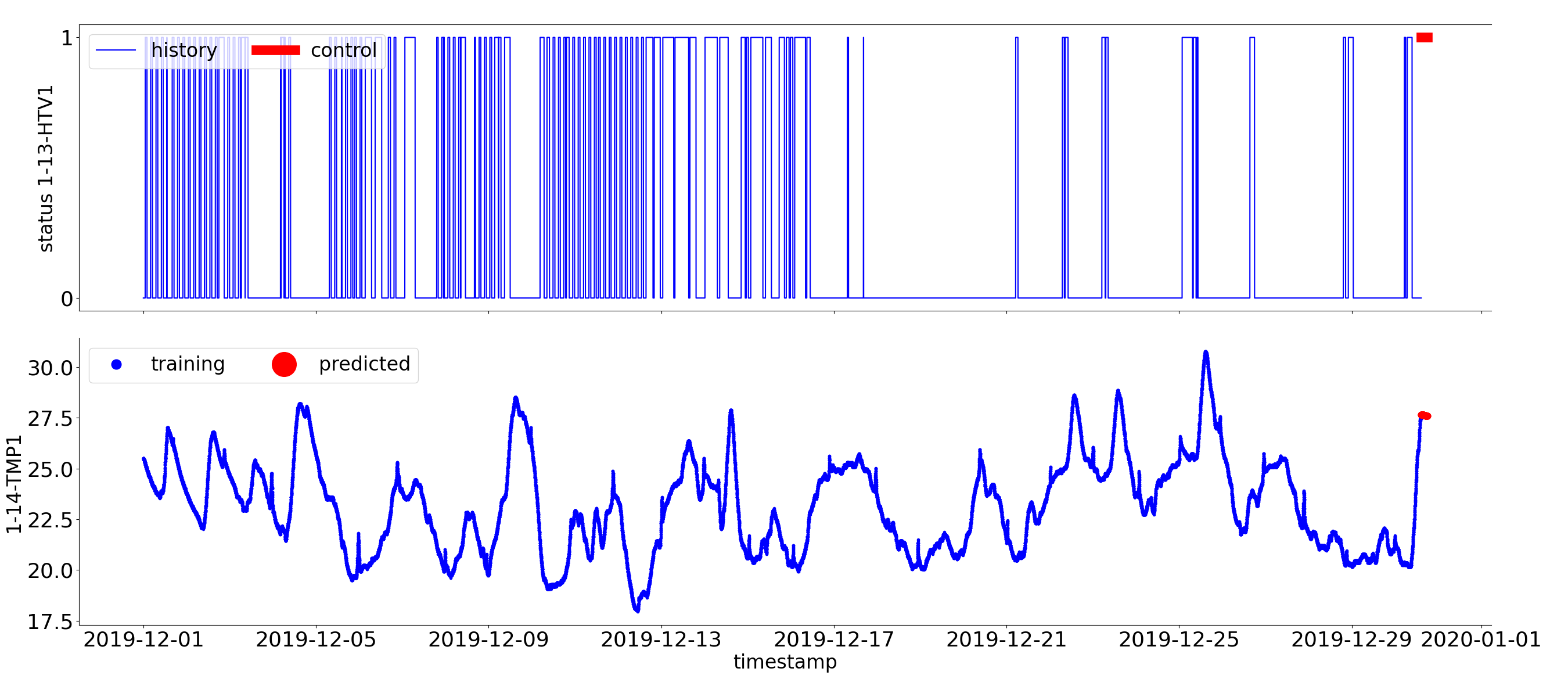}\label{fig:predict_ambient_conditions}}
\end{subfigure}
\begin{subfigure}{0.59\textwidth}
\subfloat[\footnotesize{\emph{1-14-TMP1} predictions (zoom-in of (a) on date 30/12/2019)}]{\includegraphics[clip,width=1\columnwidth,height=3cm]{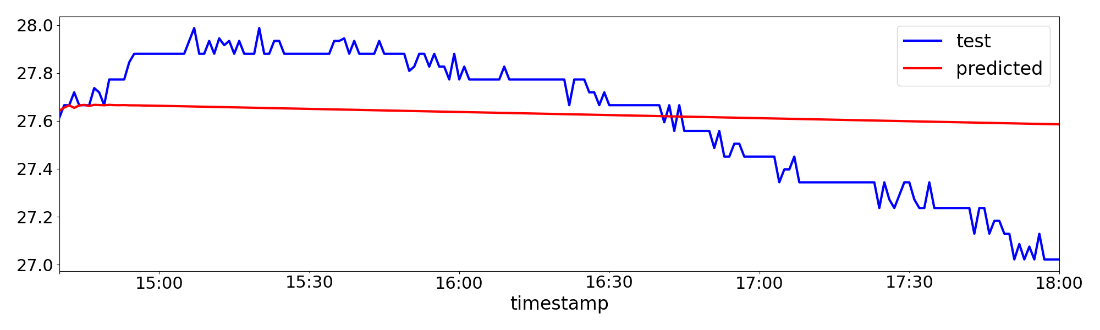}\label{fig:AR_predictions_1_14_TMP1}}
\end{subfigure}
\begin{subfigure}{0.4\textwidth}
\subfloat[\footnotesize{\emph{pacfs} \textit{vs} lags (selected: 1, 2, 3, 7)}]{\includegraphics[clip,width=1\columnwidth,height=3cm]{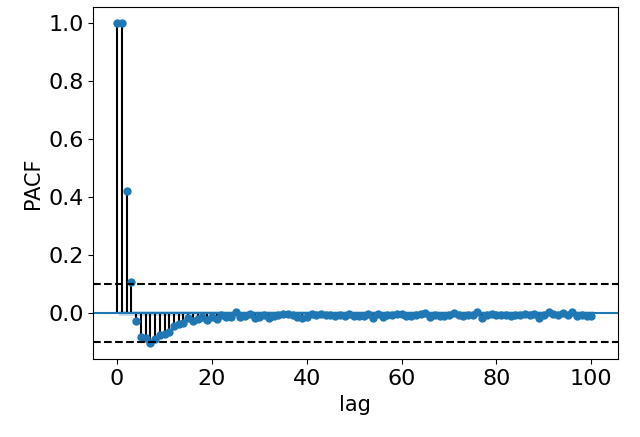}\label{fig:pacfs}}
\end{subfigure}
\caption{Ambient conditions modelling}
\label{fig:Ambient_conditions_modelling}
\end{figure}

We iterated through all predictable non-control features using original scale values. By trial and error, a uniform, absolute threshold of 0.1 was employed for partial autocorrelation function (\textit{pacf}) to identify all statistically significant lags for each AR model (Fig.\ref{fig:pacfs}). All AR models were trained with the training dataset which spans from '2019-12-01 00:00:00' to '2019-12-30 14:39:00', while the test set, which spans from ‘2019-12-30 14:40:00’ to ‘2019-12-31 23:59:00', was reserved for comparison purpose. The time span for future prediction was from ‘2019-12-30 14:40:00’ to ‘2019-12-30 18:00:00’ (totalling 201 datapoints). The future prediction scope partially overlapped with the test set horizon. They were however used for different purposes: the future prediction was used to test control variables and the control strategy, while the test set was mainly designed for evaluating model performance. In the AR model, a rolling forecasting strategy was used, in which a single prediction was made at each time step, and the prediction was appended to historical data to incrementally augment the training size. An example of the predicted trajectories versus the real test signal was shown in Fig.\ref{fig:predict_ambient_conditions} and Fig.\ref{fig:AR_predictions_1_14_TMP1}, also accompanied is the control variable 1-13-HTV1. The underfloor heating valve history roughly coincides with the indoor temperature trajectory, showing a delayed and less evident influence on 1-14-TMP1. The estimated AR model gives reasonably well predictions agreeing with the average trend due to its memory characteristic, although deviations exist, we will see later in the boosting tree case that it doesn’t affect the indoor temperature predictions to a significant level, as trees are grown in a way such that only interval values matter due to the nature of tree splitting. The control variable 1-13-HTV1 was binary and takes values 0/1 (off/on). Its value was set to 1 for all future timestamps, representing a constant on status, which is in line with our goal of warming up the room.
\vspace{3mm}
\setlength{\intextsep}{8pt}

\subsection{Indoor Temperature Modelling and Control}
In the second stage, all historical and forecasted future feature values were fed into a multivariate ML model to produce predictions for the response variable 1-15-TMP1 on the same future scale (‘2019-12-30 14:40:00’ to ‘2019-12-30 18:00:00’). Before building the ML model, a standardization procedure was applied to scale the numeric variables in Table.\ref{tab:sensor_categories}. This is useful when distance-based method (\textit{e.g.} nearest neighbors) or dimension reduction techniques are to be used. For simplicity, we only use one model, \textit{i.e.} Extreme Gradient Boosting Machine (XGBoost), to demonstrate the second stage modelling process. We first described the case where single control variable (1-13-HTV1) was used; two extra control variables, \textit{i.e.} 1-14-TMP1 and 1-8-TMP1, were introduced in a second trial. Note that, the future predictions are not comparable to the test ground truth labels, as the underlying control variable condition(s) are different.

\begin{figure}[h!]
    \centering
    \includegraphics[width=0.48\columnwidth, height=2.6cm]{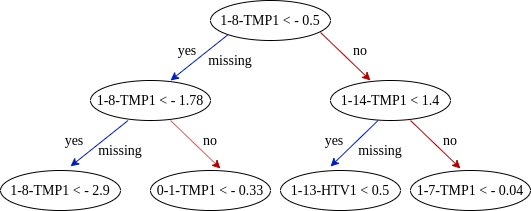}
    \caption{Depth-two diagram of the first boosted tree generated by XGBoost algorithm}
    \label{fig:xgboost_tree}
\end{figure}

A grid search was implemented to tune five model parameters, namely, number of trees, tree depth, learning rate, subsampling rates of observations and features when growing trees. The fine-tuned model used \textit{max.} 50 trees, each tree with a \textit{max.} depth of 4, learning rate of 0.3, row and column subsampling rates of 0.6 and 0.8, respectively. The trees essentially meshes the feature space and produces intervals for each variable. As an example, part of the first tree boosted is shown in Fig.\ref{fig:xgboost_tree}. The model sequentially and frequently employs 1-8-TMP1, 1-14-TMP1, 1-13-HTV1, \textit{etc}. to segment data instances, which is consistent with the feature importance ordering in Fig.\ref{fig:permutation_importance_and_correlation}. 

\vspace{2mm}
\vspace{2mm}

\begin{figure}[h!]
\centering
\begin{subfigure}{0.495\textwidth}
\subfloat[\footnotesize{case 1: control of 1-13-HTV1}]{\includegraphics[clip,width=0.85\columnwidth,height=2.5cm]{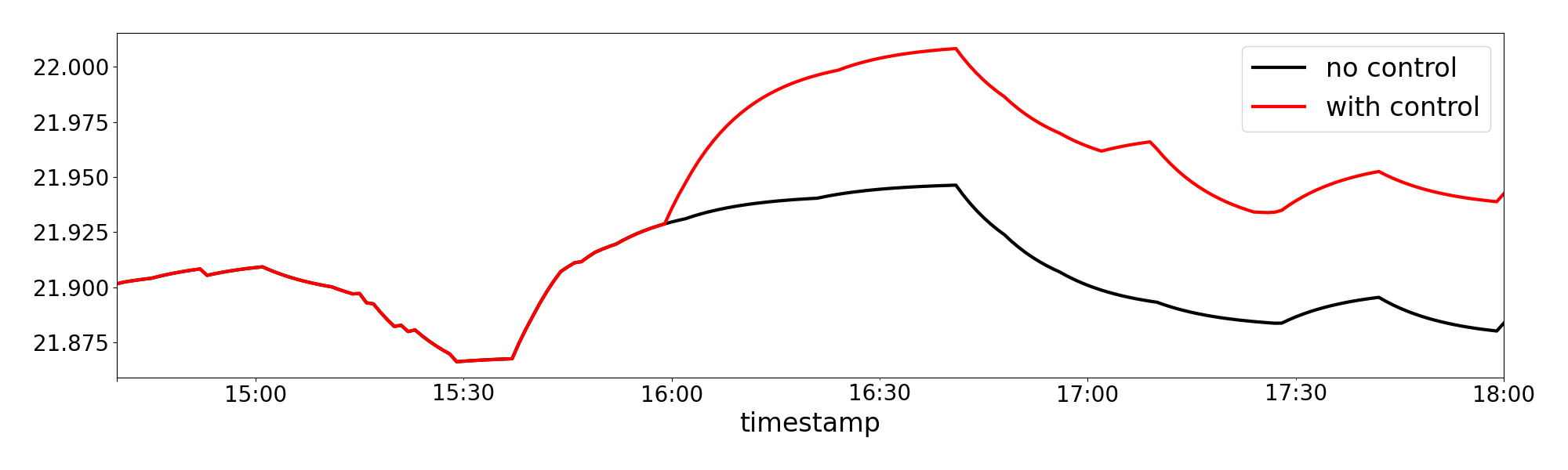}\label{fig:a}}
\end{subfigure}
\begin{subfigure}{0.495\textwidth}
\subfloat[\footnotesize{case 2: control of 1-14-TMP1 }]{\includegraphics[clip,width=0.85\columnwidth,height=2.5cm]{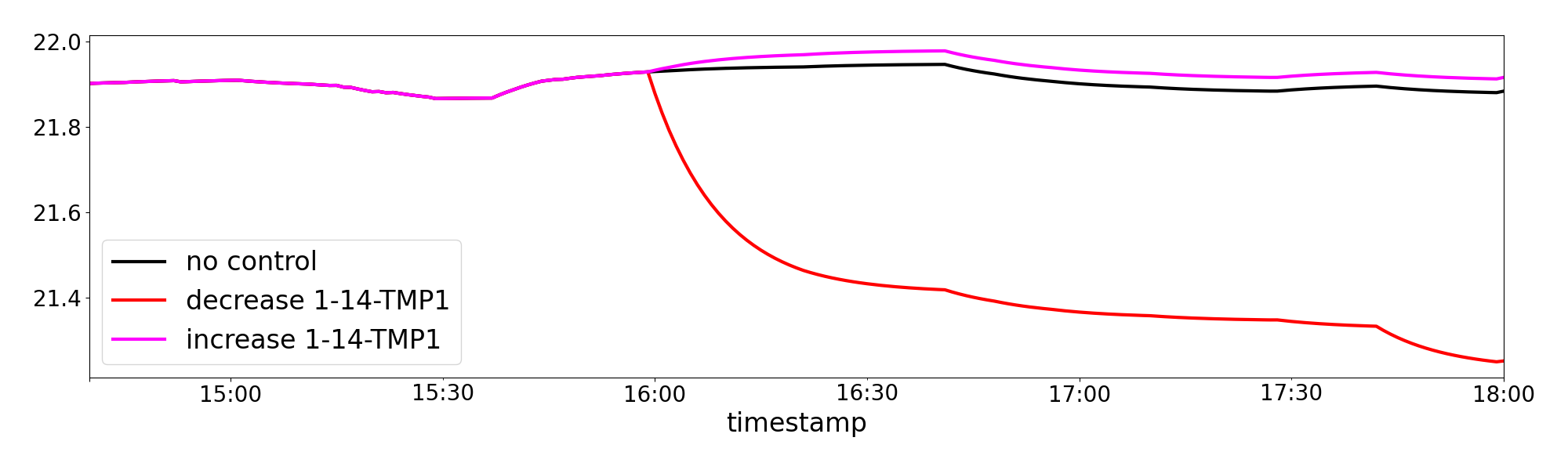}\label{fig:a}}
\end{subfigure}
\begin{subfigure}{0.495\textwidth}
\subfloat[\footnotesize{case 3: control of 1-8-TMP1}]{\includegraphics[clip,width=0.85\columnwidth,height=2.5cm]{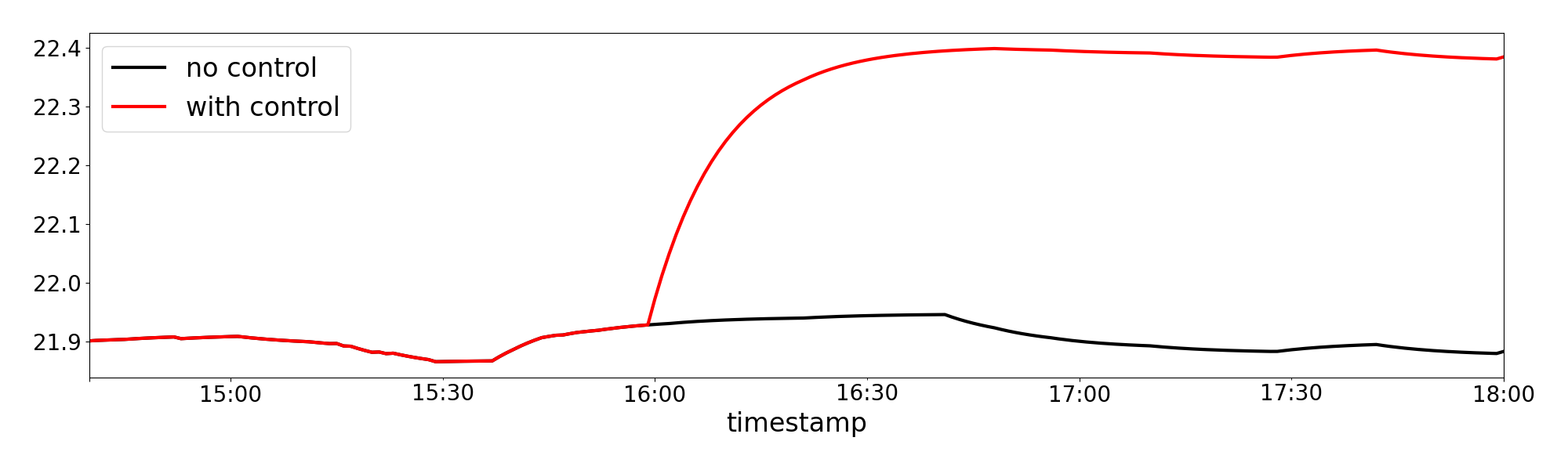}\label{fig:a}}
\end{subfigure}
\begin{subfigure}{0.495\textwidth}
\subfloat[\footnotesize{case 4: hybrid control of 1-13-HTV1, 1-14-TMP1 and 1-8-TMP1}]{\includegraphics[clip,width=0.85\columnwidth,height=2.5cm]{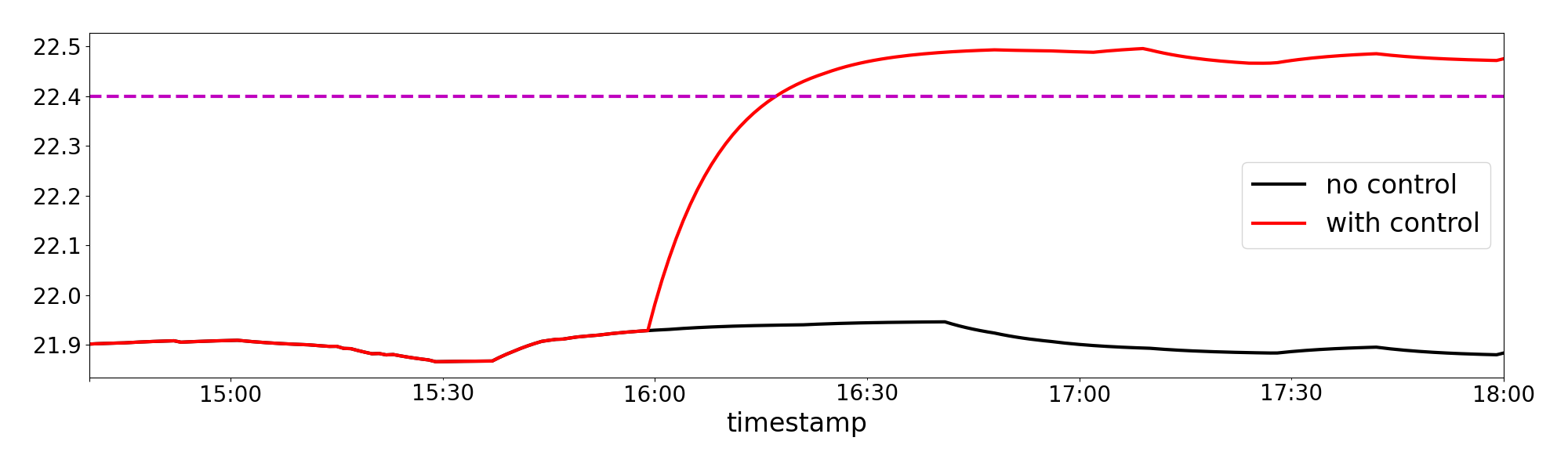}\label{fig:a}}
\end{subfigure}
\caption{Indoor temperature predictions by XGBoost, with single or hybrid controls}
\label{fig:control_experiments}
\end{figure}

On a future time scale, four control experiments were designed, as shown in Table.\ref{tab:control_experiments_design}. Three features were employed as control variables: 1-13-HTV1, 1-14-TMP1, and 1-8-TMP1. They are closely related to 1-15-TMP1, as evidenced in Fig.\ref{fig:raw_temperature_data} and Fig.\ref{fig:permutation_importance_and_correlation}, thus may give hints on inferring 1-15-TMP1. Other non-control variable values were directly obtained from the first stage AR predictions. In all the four cases, control(s) were applied starting at the time 16:00 and onwards. The experiments results were presented in Fig.\ref{fig:control_experiments}. The curves were smoothed using a 20-min moving average.

\vspace{2mm}
\setlength{\intextsep}{4pt}
\begin{table}[!htb]
  \scriptsize
  \caption{Control experiments design}
  \label{tab:control_experiments_design}
  \begin{tabular}{ccc}
    \toprule  
    \textbf{case} & \textbf{with control} & \textbf{no control} \\
    \midrule
    1 & {\color{red}1-13-HTV1 = 0/1}, 1-14-TMP1 = 26.5$^{\circ}$C, 1-8-TMP1 = 22$^{\circ}$ C & \textcolor{red}{1-13-HTV1 = 0}, 1-14-TMP1 = 26.5$^{\circ}$C, 1-8-TMP1 = 22$^{\circ}$C \\
    2 & 1-13-HTV1 = 0, \textcolor{red}{1-14-TMP1 = 26/27$^{\circ}$C}, 1-8-TMP1 = 22$^{\circ}$C & 1-13-HTV1 = 0, \textcolor{red}{1-14-TMP1 = 26.5$^{\circ}$C}, 1-8-TMP1 = 22$^{\circ}$C \\
    3 & 1-13-HTV1 = 0, 1-14-TMP1 = 26.5$^{\circ}$C, \textcolor{red}{1-8-TMP1 = 23$^{\circ}$C} & 1-13-HTV1 = 0, 1-14-TMP1 = 26.5$^{\circ}$C, \textcolor{red}{1-8-TMP1 = 22$^{\circ}$C} \\
    4 & \textcolor{red}{1-13-HTV1 = 0/1, 1-14-TMP1 = 27$^{\circ}$C, 1-8-TMP1 = 23$^{\circ}$C} & \textcolor{red}{1-13-HTV1 = 0, 1-14-TMP1 = 26.5$^{\circ}$C, 1-8-TMP1 = 22$^{\circ}$C} \\
   \bottomrule
  \end{tabular}
\end{table}
\setlength{\intextsep}{4pt}
\vspace{2.5mm}

Fig.\ref{fig:control_experiments}(a) shows the effect of control of 1-13-HTV1. The valve remained silent first (\textit{e.g.} setting the value of ‘1-13-HTV1’ to be 0), till a certain chosen timestamp (\textit{e.g.} ‘2019-12-30 16:00:00’) it was triggered. The differences between the controlled and uncontrolled scenarios are evident: generally, there is an temperature up lift of $\sim$ 0.1$^{\circ}$C after valve control was applied, which evidences that the control is effective. This is justified by the fact that 1-13-HTV1 controls the underfloor heating of the large Room 1-15 (although the valve sits in the wardrobe). Bringing in 1-14-TMP1 (temperature in the ensuite bathroom) as a control variable can also be justified: the bathroom has smaller space and it accommodates some fast heating devices (\textit{e.g.} electric radiators, heated towel rail, infrared heating panel and fan) which facilitate control. Fig.\ref{fig:control_experiments}(b) presents the results obtained by raising or decreasing the temperature in Room 1-14: cooling down the room seems to have greater effect on 1-15-TMP1 than heating it up. Fig.\ref{fig:control_experiments}(c) assumes we are able to swiftly control the temperature in Reception1 (1-8-TMP1, downstairs of 1-15-TMP1), and we can see the XGBoost model accurately captured the collaborative relation between these two variables: by increasing 1-8-TMP1 from 22$^{\circ}$C to 23$^{\circ}$C, a jump of approximately 0.4$^{\circ}$C was induced in 1-15-TMP1. This also shows 1-8-TMP1 has greater impact than 1-13-HTV1 and 1-14-TMP1 in influencing 1-15-TMP1, as has been previously observed from Fig.\ref{fig:permutation_importance_and_correlation}. The biggest impact may result from a hybrid control strategy combining all of the three, by which we expect to achieve faster, even though more aggressive heating control. The results yielded in Fig.\ref{fig:control_experiments}(d) are in line with our expectations: a jump of 0.6$^{\circ}$C was witnessed. 

The temperature changes (0.1$\sim$0.6$^{\circ}$C) in Fig.\ref{fig:control_experiments}, induced by the changes in the values of input control variables, are small but evident, implying that the XGBoost model is very sensitive in detecting variations of input controller values, and can act correspondingly to generate differentiated responses. The change is not huge as the Room 1-15 is spacious, and the influence of 1-14-TMP1 and 1-8-TMP1 are indirect. These patterns of course are obvious from a human's perspective; however, it becomes appreciable when they are identified by a learning algorithm without human intuition. 

\vspace{3mm}
\setlength{\intextsep}{4pt}

\vspace{3mm}
\subsection{Heating Control Implementation}
With the capacity of predicting future temperature trajectories, we can calculate the warm-up time needed to hit a target temperature, when the heating valve is switched on at an arbitrary timestamp. The inverse problem being, if we want the temperature to be at or above a desired value at certain timestamp, when should the heating valve be switched on? A brute-force strategy is to first predict ambient conditions over the entire period from the current time to that timestamp, and work out the closest starting point that leads to an exact match or close surpass of the target temperature at that desired timestamp. If the time gap between now and the desired point is small (during which the ambient conditions may not necessarily change much), another simpler strategy is to work out the time needed for the indoor temperature to arise to the target temperature if we switch on the valve at any future preferred time, and do an equal-length shift to the desired timestamp by assuming that the ambient conditions remain stationary or static over this time period. For simplicity, the later static approach was demonstrated using the hybrid control experiment (case 4 in Fig.\ref{fig:control_experiments}).   

We simulated a scenario to achieve heating control using the predicted indoor temperature trajectory. Imagine the following heating event was pre-set on a calendar (\textit{e.g.} pre-saved as \textit{.ical} or \textit{.yml} files):

\noindent\hspace*{2em} \emph{\small{Event:} \small{Heating}} \\
\hspace*{2em} \emph{\small{Target Zone:} \small{Room 1-15}} \\
\hspace*{2em} \emph{\small{Target Timestamp:} \small{2019-12-30 18:00:00}} \\
\hspace*{2em} \emph{\small{Target Temperature:} \small{22.40$^{\circ}$C}} \\
\hspace*{2em} \emph{\small{Controls:} \small{1-13-HTV1, 1-14-TMP1, 1-8-TMP1}} 

Suppose the current time is ‘2019-12-30 16:00:00’, so we want the temperature in Zone 1-15 to be at least, ideally around, 22.40$^{\circ}$C at 18:00 tonight. The task being, when should the heating valve be switched on to optimize utility use. The task fails if the temperature is below that at the desired time (the valve responds late); a waste of energy occurs if the target temperature is achieved earlier (the valve responds too early). The ideal case is for the temperature to arrive around the target temperature at the specified time. The complete control procedure was described in Algorithm.\ref{alg:heating_ctrol}. 
\setlength{\intextsep}{10pt}
\begin{algorithm}[t!]
{\fontsize{7.5pt}{7.5pt}\selectfont
  \caption{Learning-based heating control} 
  \label{alg:heating_ctrol}
  \begin{algorithmic}
    \Require historical data, pre-set future event
  \Statex
  \LState \textbf{Step 1: Initialization}
  
  Scan future events on calendar and obtain the future event timestamp t (‘2019-12-30 18:00:00’) and target temperature T (22.40$^{\circ}$C).
  \item[]
  \LState \textbf{Step 2: Ambient conditions forecasting}
  
        \begin{enumerate}[leftmargin=0pt,itemindent=1.5em]
            \item use a sliding window to retrieve historical data (both ambient conditions and indoor temperature).
            \item train a univariate model (\textit{e.g.} time series or \textit{LSTM}) to forecast future values for the selected features, spanning the time gap between now and $t$.
        \end{enumerate}  
  \item[]
  \LState \textbf{Step 3: Indoor temperature prediction}
  
      \begin{enumerate}[leftmargin=0pt,itemindent=1.5em]
        \item from now on, choose a before-event point $t_0$ (\textit{e.g.} now) as simulation starting point.
        \item set the control variables (single/hybrid control) to their desired values (\textit{e.g.} set valve status = on).
        \item use historical ambient conditions and indoor temperature to train a multivariate, supervised ML model (\textit{e.g.} XGBoost or neural network).
        \item plug in the forecasted future ambient conditions to predict the indoor temperature trajectory from $t_0$ to $t$.
      \end{enumerate}  
  \item[]
  \LState \textbf{Step 4: Warm-up time estimation and heating control}
  
     \begin{enumerate}[leftmargin=0pt,itemindent=1.5em]
        \item Smooth the predicted value (\emph{e.g.} using moving average), record the first timestamp $t_1$ when the predicted indoor temperature hits the target temperature $T$ and remains above for a consecutive period.
        \item calculate $\Delta t = t_1 - t_0$ as the warm-up time.
        
        If $\Delta t \geq t - now$, trigger the heating at current time. If $\Delta t < t - now$, either set the control start time as $t - \Delta t$ (static approximate control, assuming time-invariant characteristics of future ambient conditions), or go back to Step 3 and repeat the prediction-estimation process (dynamical, iterative search-based control).
        
        \item a buffer time can be further deducted from the control start time to allow extra flexibilities (sufficient warm-up time to further guarantee the target temperature is achieved).
        \item finally, set the control variables to be the desired values at the derived control start time.
      \end{enumerate}
  \item[]
\end{algorithmic}
}
\end{algorithm}
\setlength{\textfloatsep}{8pt}

Algorithm.\ref{alg:heating_ctrol} was implemented using the hybrid control strategy blending variables 1-13-HTV1, 1-14-TMP1 and 1-8-TMP1. We tentatively switch on the heating valve (1-13-HTV1=1), set 1-14-TMP1= 27$^{\circ}$C and 1-8-TMP1 =23$^{\circ}$C at ‘2019-12-30 16:00:00’, and ask the trained XGBoost model to predict how they influence the indoor temperature trajectory 1-15-TMP1. It is observed from Fig.\ref{fig:control_experiments}(d) that, given the prior temperatures 1-14-TMP1=26.5$^{\circ}$C and 1-8-TMP1=22$^{\circ}$C, the target room temperature 1-15-TMP1 was predicted to be about 21.9$^{\circ}$C initially. After heating controls were applied, the temperature started to rise and hit the target temperature 22.40$^{\circ}$C at around 16:41 and stay above it for a consecutive period of over 20 mins (this user-defined consecutively over-time defines a rule to find whether the target temperature has been achieved or not). This gives a warm-up time of 19 mins, \textit{i.e.} given the future predicted ambient conditions, if we adopt the hybrid control strategy, it takes 19 mins to arrive at the target temperature 22.40$^{\circ}$C. As the estimated warm-up time (19 mins) is less than the two-hour gap between the current time (‘2019-12-30 16:00:00’) and the desired timestamp (‘2019-12-30 18:00:00’), if ambient conditions stay approximately stationary, then we can assume the warm-up time is time-invariant, and a static approximate control strategy advises to switch on the heating valve at time ‘2019-12-30 17:41:00’; if we want to give more confidence in achieving the target temperature at the desired time, an extra 5 mins buffer time can be added, and a safe switch-on time would be ‘2019-12-30 17:36:00’. Alternatively, if variations of future ambient conditions are to be considered, this prediction-estimation process can be repeatedly exercised, starting at any arbitrary time closer to the desired timestamp to yield a potentially more accurate estimation of the warm-up time. 

Fig.\ref{fig:heating_control_implementation} recaps the complete graph of implementing the control algorithm. The AR and XGBoost models were first trained using data from ‘2019-12-01 00:00:00’ to ‘2019-12-30 14:39:00’. As stated before, even though the AR model does not give perfect accuracy, still in the second stage modelling, the performance of the XGBoost model was very impressive: given  real measured ambient conditions, the indoor temperature predictions (green and magenta dots) agree very well with ground truths (blue dots) on both the training (in-sample) and test (out-of-sample) datasets. To implement the hybrid control strategy, we manually masked the real ambient conditions data from ‘2019-12-30 14:40:00’ to ‘2019-12-30 18:00:00’ (the ‘future prediction’ scope), and asked the previously trained AR model to forecast the ambient conditions over this period, then XGBoost model was employed to further predict the indoor temperature 1-15-TMP1, with the hybrid control strategy applied. The results (red dots) provide sound advice on the most appropriate time to switch on the heating valves. Note that, the future predictions (red dots) are not directly comparable to the ground truth (blue dots), as the prior ambient conditions are different.

\vspace{2mm}
\begin{figure}[h!]
\centering
\includegraphics[clip,width=0.92\columnwidth]{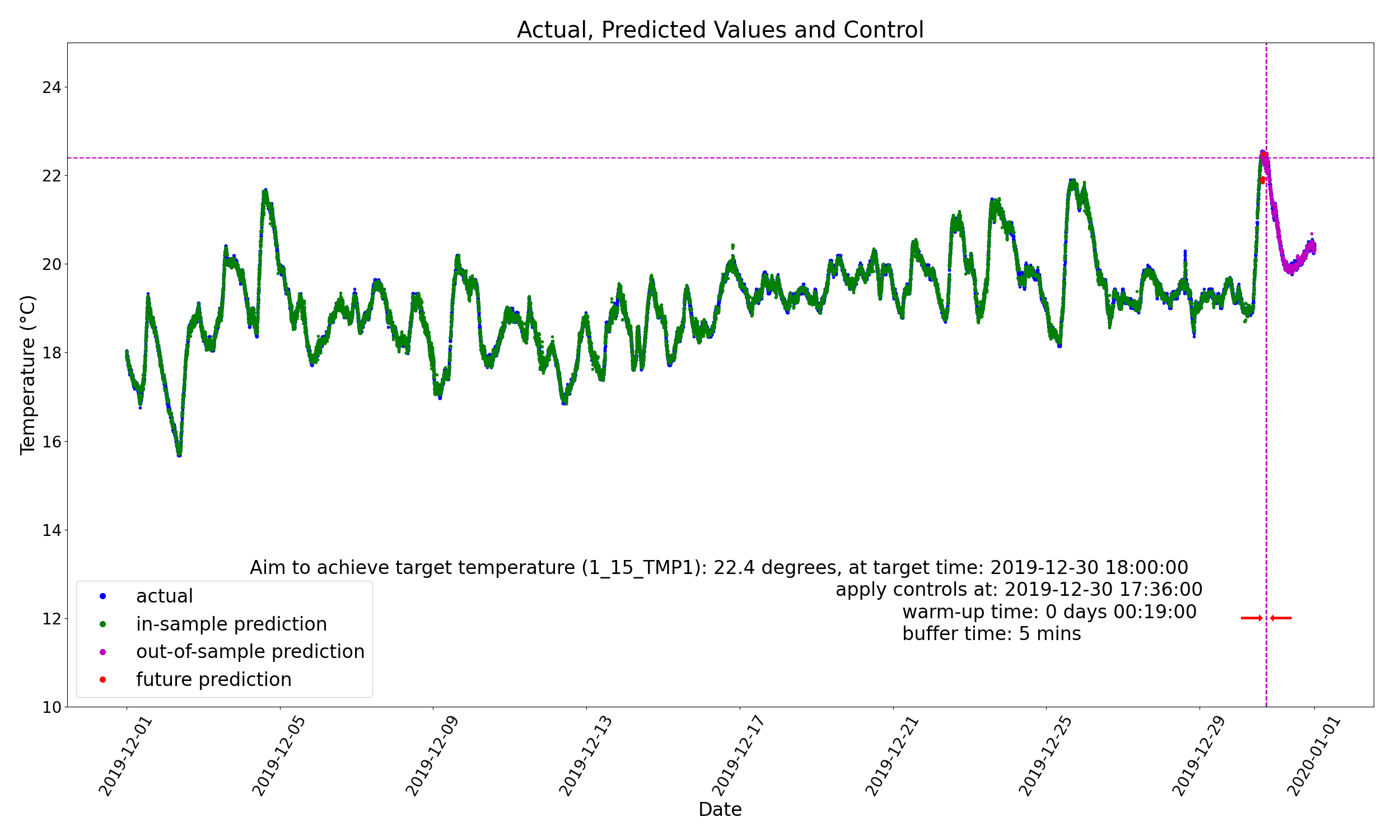}\label{fig:heating_control_implementation}
\caption{Learning-based heating control implementation}
\label{fig:heating_control_implementation}
\end{figure}

\section{Discussions}
The two-stage modelling practice is an example of empirical illustration. Once the two-stage framework is built, it can generalises to other building environments with flexible sensor inputs. In such cases, a model calibration process, along with feature selection, can be performed based on local data learning, which ensures models trained in one building can quickly adapt to new environments via pathways such as transfer learning. While AR is a mature technique in time series modelling, other potential methods could be explored. Among them, good candidates are those capable of sequential data processing, these include ARIMA, HMM and LSTM. We have researched these methods and they provide very good performance over AR; however, they also consume more computing resources and pose more requirements on real-time computing. The performance of the second stage learning, \textit{i.e.} the XGBoost, was impressive. This is built on the success of feature forecasting in the first stage. Both stages have different levels of error tolerance, due to the nature of tree-based models (weak learners are ensembled sequentially into a strong learner), good performance was achieved. It would be good practice to compare this two-stage modelling approach with other modelling techniques such as vector AR (VAR) which directly predicts the future using multiple time series inputs.

The control strategy developed answers the question of when to turn on the control variables; however, it can be used to decide when to turn off as well, by repeating the same algorithm in the post-control stage (\textit{i.e.} after the target temperature has been achieved), intervally turning of some of the control variables and checking the resulting future temperature trajectories. So it’s a solution for both tasks. An intelligent BMS can automatically manage different operation variables such as radiators, lights, water tanks, air conditioners, and even appliance. Theoretically, adding in more positively correlated (with the response variable) control variables could benefit the control strategy, and accelerate the heating process. This was demonstrated in the case where the heating valve 1-13-HTV1, the next door temperature 1-14-TMP1, and downstairs  temperature 1-8-TMP1, join together to form a hybrid control strategy. However, more investigations are needed to study the join effect of mixed control variables.

The significance of this work lies in following aspects: first, it builds up an automated, AI-powered, data-driven forecasting system for heating control, this framework contributes to efficient energy utilisation in line with user request, and improve sustainability in smart buildings. It learns from environments (ambient conditions) and the learned patterns makes this system ‘well-behaved’ (saving energy while retaining comfort level) in a real-time environment. It’s more flexible compared to traditional rule-based control systems, also more efficient than conventional thermal dynamics-based mathematical models as it evolves dynamically, reducing the labor of reconstructing structured models. Secondly, this work combines time series modelling and ML to form a two-stage sequential modelling approach. It allows some flexibilities for control (via the set of control variables) as well as dynamical learning from environments (ambient conditions). Time series model captures the temporal dependence in each feature, while ML model reveals the cross-sectional correlations between features and response. This mixed temporal-spatial approach enables real time control of the quantity concerned.

\section{Conclusions}

This research introduces a sequentially pipelined, mixed data-driven approach to modelling building data, from which indoor temperature patterns can be learned and optimal heating control decisions can be made. Sensory data was collected through a wireless sensor network, information were fed into an intelligent building management system to be analysed and to advise control actions. The two-stage modelling framework sequentially connects time series and machine learning methods, and combines learning and input control. A heating control algorithm was designed to maximize energy utilization, and the effectiveness of the single and hybrid control strategies was demonstrated. This work contributes towards scalable, automated, and energy-efficient intelligent building management system design and effective service control to meet human comfort expectation.

\bibliographystyle{unsrt}
\bibliography{reference}

\end{document}